\documentclass[prd,amsmath,amssymb]{revtex4}

\usepackage{graphicx}% Include figure files
\usepackage{bm}% bold math

\normalbaselines \topmargin -0.25 truein \oddsidemargin 0.30
truein \evensidemargin 0.30 truein 
\begin{document}

\title{Axionic extension of the Einstein-Dirac-aether theory: \\ Degeneracy removal with respect to shear of the aether flow}

\author{Anna O. Efremova}
\email{anna.efremova131@yandex.ru} \affiliation{Department of General
Relativity and Gravitation, Institute of Physics, Kazan Federal University, Kremlevskaya
str. 18, Kazan 420008, Russia}
\author{Alexander B. Balakin}
\email{Alexander.Balakin@kpfu.ru} \affiliation{Department of General
Relativity and Gravitation, Institute of Physics, Kazan Federal University, Kremlevskaya
str. 18, Kazan 420008, Russia}

\date{\today}

\begin{abstract}
We consider dynamics of the quartet of interacting cosmic substrata, which includes the dynamic aether, presented by the unit timelike vector field, the axionic dark matter, described by the pseudoscalar field, the spinor field associated with fermion particles, and the gravity field.  The extended set of master equations is derived based on the idea that the potential of the axion field to be the function of seven arguments. The first one is, standardly, the pseudoscalar field; the second and third arguments are the fundamental spinor invariant and pseudoinvariant; the fourth and fifth ones are the aether-axion cross-invariants and cross-pseudoinvariants; the sixth argument is the expansion scalar, and the seventh one is the square of tensor of the shear of the aether flow. The complete set of master equations is derived and prepared for analysis.
\end{abstract}

\maketitle

\section{Introduction}\label{Intro}

We study the guiding role of dynamic aether in the processes of interaction between spinor and pseudoscalar fields. This sector of cosmic interactions attracts attention, when one considers the coupling between the axionic dark matter and fermions (massive Dirac particles), as well as, the coupling between axions and massless neutrinos \cite{1}. Since the dynamic aether is mathematically associated with the global timelike unit vector field $U^j$, the effect of this cosmic substratum on the axion and spinor field could be associated with the acceleration four-vector $a_i = U^j \nabla_j U_i$, with the expansion scalar $\Theta = \nabla_k U^k$, with the skew-symmetric vorticity tensor $\omega_{mn}$ and the symmetric traceless shear tensor $\sigma_{mn}$. When we work with the isotropic Friedmann type cosmology, the only $\Theta$ turns out to be not equal to zero. The nonvanishing vorticity tensor appears in the G\"odel's rotating Universe. When we deal with the Bianchi type anisotropic homogeneous cosmological models, the nonvanishing shear tensor inevitably appears as an important characteristic of the dynamic aether flow. Five years ago we were faced with specific degeneracy with respect to the shear of the aether velocity. In 2017, the observation of the binary neutron star merger \cite{170817} has shown that the ratio of the velocities of the gravitational and electromagnetic waves is highly likely equal to one, so that two Jacobson's constants $C_1$ and $C_3$ \cite{J} are linked by the relationship $C_1{+}C_3{=}0$. This means that the shear tensor happens to be crossed out from the aether kinetic term, thus disappearing from the Lagrangian of the dynamic aether and becoming the hidden quantity.

We suggest the mechanism of the degeneracy removal with respect to the shear of the aether flow, which is connected with the introduction of the square of the shear tensor $\sigma^2 \equiv \sigma_{mn} \sigma^{mn}$ into the potential of the axion field $V$. Also, we assume that  the potential $V$ is the function of $\Theta$ and depends on the spinor scalar $S=\bar{\psi} \psi$, on the spinor pseudoscalar $S=\bar{\psi} \gamma^5\psi$, as well as, on the cross-invariant $\omega= U_n\bar{\psi} \gamma^n \psi$ and cross-pseudoinvariant   $\Omega=U_n \bar{\psi} \gamma^n \gamma^5\psi$. (Here and below $\gamma^n$ is the Dirac matrices). As the result, master equations of all the presented fields are shown to acquire new specific terms; we derive them and reduce to the case of the Bianchi-I model.

\section{The formalism}

The action functional of the Einstein-Dirac-aether-axion theory contains three groups of terms:
\begin{equation}
S =  \int d^4 x \sqrt{{-}g} \ \left\{ \frac{1}{2\kappa}\left[R{+}2\Lambda {+} \lambda (g_{mn}U^m
U^n {-}1 ){+} K^{abmn} \nabla_a U_m \nabla_b U_n \right] {+} \right.
\label{1}
\end{equation}
$$
\left. + \left[\frac{i}{2}\left(\bar{\psi}\gamma^{k}D_{k} \psi {-}D_{k}\bar{\psi}\gamma^{k}\psi\right) {-}  m \bar{\psi}\psi \right]
+ \frac{1}{2}\Psi^2_0 \left[V {-} g^{mn}\nabla_m \phi \nabla_n \phi \right] \right\}
\,.
$$
The first group located in square brackets with the multiplier $\frac{1}{2\kappa}$ in front, relates to the version of the vector-tensor theory of gravity known as the Einstein-aether theory \cite{J}. In this theory the variation with respect to the Lagrange multiplier $\lambda$ provides the aether velocity four-vector $U^k$ to be unit, $g_{mn}U^m U^n {=} 1$. The kinetic term
${\cal K} \equiv K^{abmn} \nabla_a U_m \nabla_b U_n $ with the constitutive tensor
\begin{equation}
K^{abmn}{=} C_1 g^{ab} g^{mn} {+} C_2 g^{am} g^{bn}
{+} C_3 g^{an} g^{bm} {+} C_4 U^{a} U^{b}g^{mn} \,,
\label{K}
\end{equation}
in which we have to put $C_3=-C_1$ due to results of the observations \cite{170817}, can be rewritten as
\begin{equation}
{\cal K} =(C_1 {+} C_4)DU_k DU^k {+}
2C_1 \omega_{ik} \omega^{ik} {+} C_2  \Theta^2
\,. \label{act5n}
\end{equation}
Here we use the  decomposition of the covariant derivative of the aether velocity four-vector
\begin{equation}
\nabla_iU_k = U_i DU_k {+} \sigma_{ik} {+} \omega_{ik}{+} \frac13 \Theta \Delta_{ik} \,, \quad D = U^j\nabla_j \,, \quad \Delta_{ik}=g_{ik}{-}U_iU_k \,,
\label{decompos}
\end{equation}
where the symmetric traceless shear tensor and the vorticity tensor have, respectively, the form
\begin{equation}
\sigma_{ik}= \frac12\Delta^p_{i}\Delta^q_{k}\left(\nabla_pU_q {+} \nabla_qU_p)\right) {-}\frac13 \Theta \Delta_{ik} \,, \quad \omega_{ik}=\frac12 \Delta^p_{i}\Delta^q_{k}(\nabla_pU_q {-} \nabla_qU_p) \,.
\label{decompos2}
\end{equation}
The second group of terms relates to the contribution of the Dirac field. It includes the spinor field $\psi$ and its Dirac conjugated $\bar{\psi}$, the Dirac matrices $\gamma^n$ and $\gamma^5$. The covariant derivatives of the spinor field are taken in the Fock-Ivanenko form based on the tetrad four-vectors $X^j_{(a)}$:
\begin{equation}
D_{k} \psi =  \partial_k \psi - \Gamma_k \psi   \,, \quad D_{k}\bar{\psi} = \partial_k \bar{\psi} + \bar{\psi} \Gamma_k  \,, \quad \Gamma_{k}=\frac{1}{4}g_{mn}X^{(a)}_{s}\gamma^{s}\gamma^{n} \nabla_{k}X^{m}_{(a)}  \,.
\label{decompos4}
\end{equation}
The third group of terms with the square of the coupling constant $\Psi_0$ in front, describes the contribution of the pseudoscalar (axion) field $\phi$.
The potential of the axion field is considered to be the function of seven arguments (see their definitions above) $V = V\left(\phi, S, P, \omega, \Omega, \Theta, \sigma^2 \right)$.
The arguments $S$ and $P$ describe the specific coupling of the axion and spinor fields; the arguments $\omega$ and $\Omega$ describe trilateral interactions between axion, spinor fields and aether; the arguments $\Theta$ and $\sigma^2$ describes the direct control over the axion field, which is carried out by the aether.

{\it Master equation for the axion field}, obtained  by the variation of the action functional (\ref{1}) with respect to $\phi$, keeps the standard form:
\begin{equation}
g^{mn} \nabla_m \nabla_n \phi = - \frac12 \frac{\partial }{\partial \phi}V(\phi, S,P,\omega, \Omega,\Theta, \sigma^2) \,.
\label{ax10}
\end{equation}
{\it Dirac equations} acquire new terms due to variation of the potential $V$ with respect to $\bar{\psi}$ and $\psi$:
\begin{equation}
i\gamma^{n} D_{n}\psi - \left(m E {-} \frac12 \Psi^2_0 \tilde{M} \right) \psi =0 \,, \quad i D_{n}\bar\psi\gamma^{n} + \bar\psi \left(m E {-} \frac12 \Psi^2_0 \tilde{M} \right)  = 0 \,,
\label{Dirac}
\end{equation}
\begin{equation}
\tilde{M} {=}  \frac{\partial V}{\partial S} E  {+} \frac{\partial V}{\partial P} \gamma^5 {+} \left(\frac{\partial V}{\partial \omega} E  {+}
\frac{\partial L}{\partial \Omega} \gamma^5 \right) U_k \gamma^k {+}
\frac{\partial V}{\partial S} E  {+} \frac{\partial V}{\partial P} \gamma^5 {+} \left(\frac{\partial V}{\partial \omega} E  {+}
\frac{\partial V}{\partial \Omega} \gamma^5 \right) U_k \gamma^k  \,.
\label{M}
\end{equation}
The matrix term $M = \left(m E {-} \frac12 \Psi^2_0 \tilde{M} \right)$ plays the role of the effective mass of the spinor field interacting with the axion field; $E$ is the unit matrix (see \cite{1} for details).

{\it Master equations for the unit vector field} $U^j$  undergo the following extensions:
\begin{equation}
\nabla_a {\cal J}^{aj} = \lambda \ U^j  + I^j \,, \quad \lambda =  U_j \left[\nabla_a {\cal J}^{aj}- I^j \right]  \,,
\label{0A1}
\end{equation}
\begin{equation}
I^j =  C_4 (DU_m)(\nabla^j U^m) + \kappa \Psi_0^2 \left[\frac {\partial V}{\partial \omega}(\bar{\psi} \gamma^j \psi)  + \frac {\partial V}{\partial \Omega}(\bar{\psi} \gamma^j \gamma^5 \psi) - \frac {\partial V}{\partial (\sigma^2)} \sigma^{jm} DU_m \right]\,,
\label{0A22}
\end{equation}
\begin{equation}
{\cal J}^{aj} = {\cal J}^{(0)aj}  + \kappa \Psi_0^2 \left [\frac {\partial V}{\partial (\sigma^2)} \sigma^{aj}+\frac12 \frac {\partial V}{\partial \Theta} g^{aj}\right] \,, \quad
{\cal J}^{(0)aj} = {K}^{abjn} (\nabla_b U_n) \,.
\label{0A33}
\end{equation}
{\it Equations for the gravity field} as the result of variation with respect to metric is of the form
\begin{equation}
R_{ik} - \frac{1}{2} R \ g_{ik}
=  \Lambda g_{ik}  + \kappa T^{(D)}_{ik} + \kappa T^{({\rm A})}_{ik} +  T_{ik}^{(U)} + T_{ik}^{(V)}  \,. \label{0Ein1}
\end{equation}
The canonic stress-energy tensors of the spinor and pure axion field have the standard form
\begin{equation}
T^{(\rm D)}_{ik} = \frac{i}{2}\left[\bar\psi \gamma_{(i} D_{k)}\psi {-} D_{(i}\bar\psi \gamma_{k)} \psi \right] {-} g_{ik} \left\{\frac{i}{2}\left[\bar\psi \gamma^{n} D_{n}\psi {-} (D_{n}\bar\psi) \gamma^{n} \psi \right] {-} m \bar\psi \psi \right\} \,,
\label{TD}
\end{equation}
\begin{equation}
T^{(A)}_{ik} = \Psi^2_0 \left[\nabla_i \phi \nabla_k \phi
+\frac12 g_{ik}\left(V {-} \nabla_n \phi \nabla^n \phi \right) \right] \,.
\label{qq1}
\end{equation}
The term $T_{ik}^{(U)}$ contains the derivatives of the axion field potential with respect to $\Theta$ and $\sigma^2$:
$$
T_{ik}^{(U)} =
\frac12 g_{ik} \ K^{abmn} \nabla_a U_m \nabla_b U_n{+} U_iU_k U_j \nabla_a {\cal J}^{aj} {+} C_4 \left(D U_i D U_k {-} U_iU_k DU_m DU^m \right) {+}
$$
\begin{equation}
{+}\nabla^m \left[U_{(i}{\cal J}^{(0)}_{k)m} {-}
{\cal J}^{(0)}_{m(i}U_{k)} {-}
{\cal J}^{(0)}_{(ik)} U_m\right]{+}
C_1\left[(\nabla_mU_i)(\nabla^m U_k) {-} (\nabla_i U_m )(\nabla_k U^m) \right]
\,,
\label{5Ein1}
\end{equation}
The term $T_{ik}^{(V)}$, in addition, contains the derivatives of $V$ with respect to $\omega$ and $\Omega$:
\begin{equation}
T_{ik}^{(V)} = 2\kappa \Psi_0^2 \left \{ \frac {\partial V}{\partial \omega}\left[\bar{\psi} U_{(i} \gamma_{k)}\psi \right] {+}  \frac {\partial V}{\partial \Omega}\left[\bar{\psi} U_{(i} \gamma_{k)} \gamma^5 \psi \right]{+} \right.
\label{VV}
\end{equation}
$$
\left.
+  \frac {\partial V}{\partial (\sigma^2)} \left[\sigma_{m(i}\sigma_{k)}^{m} {-} \sigma_{m(i}\nabla_{k)}U^m {+} \frac13 \Theta \sigma_{ik}\right] {-} \frac12 (D{+}\Theta)\left [\sigma_{ik} \frac {\partial V}{\partial (\sigma^2)}  {+} \frac12 g_{ik} \frac {\partial V}{\partial \Theta} \right]\right \} \,.
$$
The complete system of master equations for gravitational, spinor, vector and axion fields is derived and is ready for analysis; unfortunately, it is out of the frames of this short note.

\section*{Outlook}

We plan to apply the formulated theory to the anisotropic homogeneous cosmological model of the Bianchi-I type, for which $\sigma_{ik} \neq 0$.  For this case the structure of the potential of the axion field $V(\phi, S, P, \omega, \Omega, \Theta, \sigma^2)$ guarantees the degeneracy removal with respect to the shear tensor, attributed to the aether velocity.

%%%%%%%%%%%%%%%%%%%%%%%%%%%%%%%%%%%%%%%%%%%%%%%%%%%%%%%%%%%

\end{document}